\documentclass[english,aps,prl,twocolumn,prl,showpacs]{revtex4}
\usepackage[T1]{fontenc}
\usepackage[latin9]{inputenc}
\usepackage{color}
\usepackage{verbatim}
\usepackage{amssymb}
\usepackage{graphicx}
\usepackage{esint}

\makeatletter
\@ifundefined{textcolor}{}
{%
 \definecolor{BLACK}{gray}{0}
 \definecolor{WHITE}{gray}{1}
 \definecolor{RED}{rgb}{1,0,0}
 \definecolor{GREEN}{rgb}{0,1,0}
 \definecolor{BLUE}{rgb}{0,0,1}
 \definecolor{CYAN}{cmyk}{1,0,0,0}
 \definecolor{MAGENTA}{cmyk}{0,1,0,0}
 \definecolor{YELLOW}{cmyk}{0,0,1,0}
}

\makeatother

\usepackage{babel}
\begin{document}

\title{Decoherence of Impurities in a Fermi Sea of Ultracold Atoms}

\author{Marko Cetina$^{1,2}$, Michael Jag$^{1,2}$, Rianne S. Lous$^{1,2}$,
Jook T. M. Walraven$^{1,3}$, Rudolf Grimm$^{1,2}$}

\affiliation{$^{1}$Institut f\"{u}r Quantenoptik und Quanteninformation (IQOQI),
\"{O}sterreichische Akademie der Wissenschaften, 6020 Innsbruck, Austria}

\affiliation{$^{2}$Institut f\"{u}r Experimentalphysik, Universit\"{a}t Innsbruck, 6020 
Innsbruck, Austria}

\affiliation{$^{3}$Van der Waals-Zeeman Institute, Institute of Physics, University of Amsterdam, Science Park 904, 1098 XH Amsterdam, The Netherlands}

\author{Rasmus S. Christensen$^{4}$, Georg M. Bruun$^{4}$}

\affiliation{$^{4}$Department of Physics and Astronomy, Aarhus University, DK-8000 Aarhus C, Denmark}

\date{\today}

\pacs{67.85.Lm, 03.75.Dg, 03.65.Yz, 34.50.Cx, 71.38.-k}
\begin{abstract}
We investigate the decoherence of $^{40}$K impurities interacting
with a three-dimensional Fermi sea of $^{6}$Li across an interspecies
Feshbach resonance. The decoherence is measured as a function of the interaction strength and temperature using a spin-echo atom interferometry
method. For weak to moderate interaction strengths, we interpret our measurements in terms of scattering of K quasiparticles by the Fermi sea 
and find very good agreement with a Fermi liquid calculation.
For strong interactions, 
we observe significant enhancement of the decoherence rate, which 
is largely independent of temperature, pointing to behavior that is 
beyond the Fermi liquid picture.
\end{abstract}
\maketitle
Many-body fermionic systems with strong interactions play a central
role in condensed-matter, nuclear, and high-energy physics. The intricate quantum 
correlations between fermions challenge our understanding of these 
systems. Mixtures of ultracold fermionic gases offer outstanding opportunities 
to study strongly interacting fermions experimentally. Since the turn 
of the century, the excellent control over the strength of the interaction 
and the composition of these mixtures has allowed investigations addressing 
the broad spectrum from few-body to many-body phenomena \cite{Bloch2008mbp,Giorgini2008tou}. 
Tuning of the interaction is achieved using Feshbach resonances \cite{Chin2010fri}. 
The composition is varied by selecting internal states or by mixing 
different atomic species. This development has led to many exciting 
results concerning the quantum phases of fermionic mixtures, their excitations, 
superfluid behavior, and the equation of state \cite{Bennemann2015nsv2}.

In two-component fermionic systems with a large population imbalance,
the minority atoms have been shown to form quasiparticles 
termed Fermi polarons, even for surprisingly large coupling strengths~\cite{Schirotzek2009oof,Kohstall2012mac,Koschorreck2012aar,Massignan2014pdm}.
These are long-lived states described by Fermi liquid theory \cite{Baym1991lfl}.
Their lifetime is limited by scattering against the majority atoms, which is suppressed by Pauli blocking as the 
temperature approaches zero \cite{Landau1957oia,Landau1957tof}. Although 
the quasiparticle scattering rate has been determined in two-dimensional 
electron gases \cite{Berk1995rmo,Murphy1995lot,Slutsky1996ees}, measurements 
in well-defined three-dimensional (3D) fermionic systems have remained
an experimental challenge.

Intriguing questions are related to the behavior of impurities and,
more generally, Fermi mixtures in the strongly interacting regime
\cite{Massignan2014pdm,Nascimbene2011flb,Sagi2015bot}. For investigating
an impurity in a Fermi sea, Ref.\ \cite{Knap2012tdi} suggested 
a time-domain method that is applicable for a wide range of interaction
strengths. This approach can be regarded as a measurement of the coherence
of a superposition of internal states of the impurity atoms using
 interferometery \cite{Cronin2009oai}. Atom coherence has previously
been used to probe many-body demagnetization in fermionic systems
\cite{Bardon2014tdd} and impurity scattering in bosonic systems \cite{Scelle2014mco}. 

In this Letter, we report on measurements of decoherence of K
atoms immersed in a Fermi sea of Li using the method proposed
in Ref.\ \cite{Knap2012tdi}, in the regime of strong population
imbalance. We tune the interaction between the Li and K atoms using
an interspecies Feshbach resonance (FR). 
For weak to moderately strong interactions, 
we interpret the measured decoherence in terms of scattering of K 
quasiparticles by the Li Fermi sea. We find very good agreement with a Fermi liquid calculation.
This provides a 
determination of the quasiparticle scattering rate in a clean 
3D fermionic system. We extend our measurements to strong Li-K 
interactions and find decoherence rates that are almost 
an order of magnitude faster than can be explained by quasiparticle 
scattering. These decoherence rates do not increase with temperature, 
which is an indication of zero-temperature quantum dynamics in a fermionic 
many-body system.

The starting point of our experiments is an evaporatively cooled,
thermally equilibrated mixture of typically 3$\times$$10^{5}$ $^{6}$Li
atoms%
and 1.5$\times$$10^{4}$ $^{40}$K atoms%
, trapped in a crossed-beam 1064-nm optical dipole trap under conditions
similar to those in Ref.\ \cite{Kohstall2012mac}. The Li cloud is
degenerate, with $k_{B}T$/$\epsilon_{F}$ as low as 0.15%
, where $T$ is the temperature and $\epsilon_{F}$$\approx$$h$$\times$35 
kHz%
{} is the average Li Fermi energy sampled by the K atoms. Because of
the Li Fermi pressure and the more than two times stronger optical
potential for K, the K cloud is much smaller than the Li cloud \cite{Trenkwalder2011heo},
and therefore samples a nearly homogenous Li environment, with a standard
deviation in the local Li Fermi energy of less than $0.1\,\epsilon_{F}$%
. In spite of the smaller size of the K cloud, the concentration of
K in the Li sea remains low, with $\bar{n}_{{\rm K}}/\bar{n}_{{\rm Li}}$$\approx$0.3%
, where $\bar{n}_{{\rm K}}$ ($\bar{n}_{{\rm Li}}$) is the average
K (Li) number density sampled by the K atoms. The K ensemble is correspondingly
non-degenerate, with $k_{B}T/E_{F}^{{\rm K}}>0.9$, where $E_{F}^{{\rm K}}$
is the peak K Fermi energy.

We tune the interaction between the K and Li atoms
using an interspecies FR between the Li atoms in the lowest Zeeman
sub-level Li$\left|1\right\rangle $ and K atoms 
in the third-lowest sub-level K$|3\rangle$ 
\cite{Naik2011fri}. We quantify the interactions between Li and K 
by the dimensionless interaction parameter $-1/\kappa_{F}a$, where 
$\kappa_{F}$=$\hbar^{-1}\sqrt{2m_{{\rm Li}}\epsilon_{F}}$ is the Li 
Fermi wavenumber with $m_{{\rm Li}}$ the Li mass, and $a$ 
is the $s$-wave interspecies scattering length. The latter can be 
tuned as $a$=$a_{{\rm bg}}[1$$-$$\Delta/(B$$-$$B_{0})]$ by applying a magnetic
field $B$, where $B_{0}$$\approx$154.7 G is the resonance center,
$a_{{\rm bg}}$=63.0 $a_{0}$ ($a_{0}$ is Bohr's radius) and $\Delta$=880 mG
\cite{Naik2011fri}. The relatively narrow nature of our FR causes
significant momentum dependence of the interspecies interaction. We
characterize this effect by the length parameter $R^{*}$ 
\cite{Petrov2004tbp, Kohstall2012mac}. In our
experiments $\kappa_{F}R^{*}$ is approximately 0.9%
, corresponding to an intermediate regime where the interaction is near-universal with substantial effective-range 
effects.

\begin{figure}[t]
\includegraphics[scale=0.45]{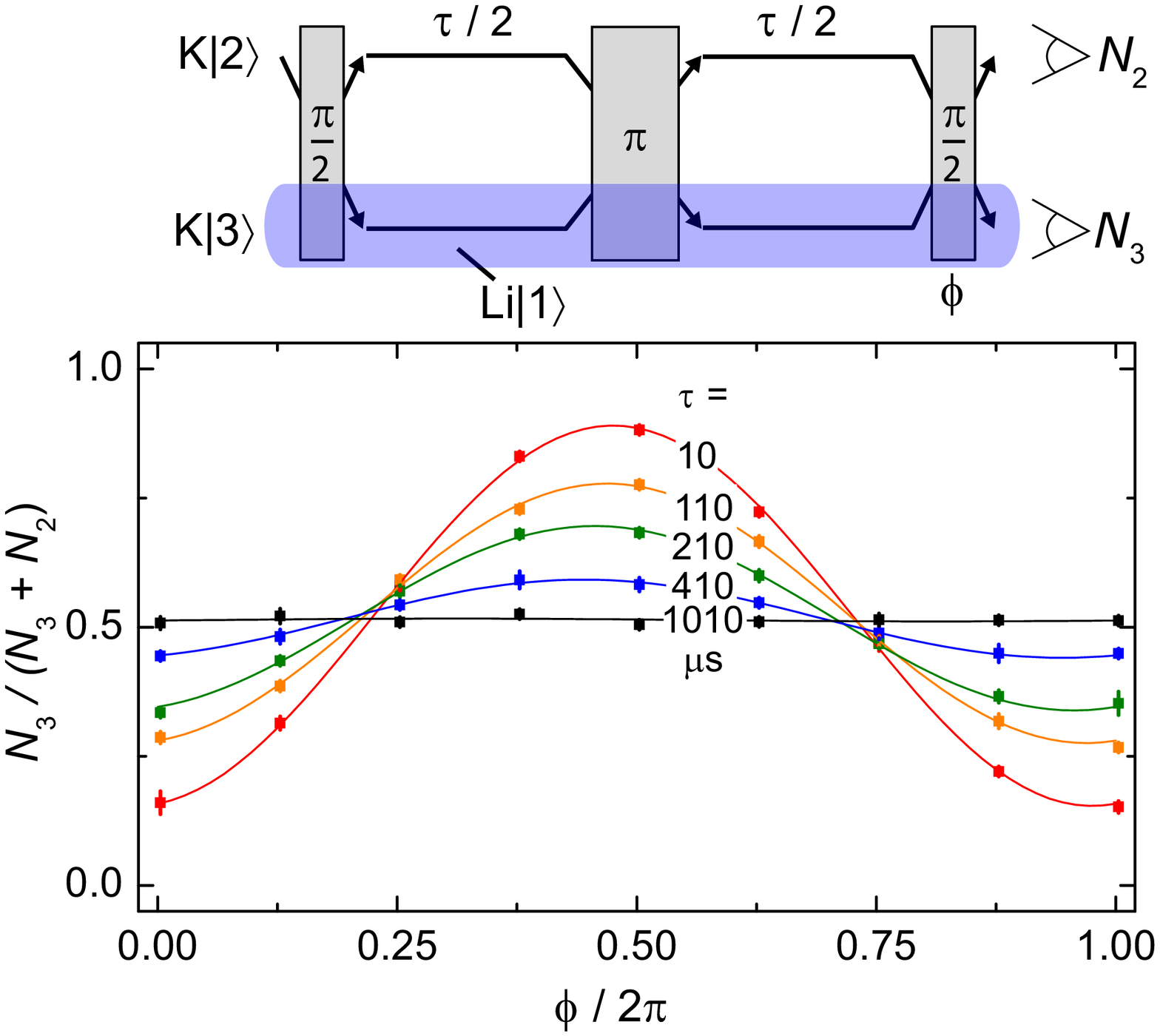}\caption{\label{fig:1}Interferometeric method for measuring the decoherence
of K in a Li Fermi sea. The upper illustration shows
a schematic of the rf pulse sequence. The atoms in the K$\left|3\right\rangle $
state interact with a Fermi sea of Li$\left|1\right\rangle $ atoms,
as indicated by the shaded region. The graph shows the fraction of
the K atoms transferred to the K$\left|3\right\rangle $ state as
a function of the relative phase of the final $\pi/2$ rf pulse for
various interaction times $\tau$ and for $-1/\kappa_{F}a$=1.9, 
$T$=0.16 $\epsilon_{F}/k_{B}$.}
\end{figure}

We probe the decoherence of the K atoms using a radio-frequency (rf)
interferometric technique, as illustrated in Fig.\ \ref{fig:1}.
The K atoms are initially prepared in the second-lowest Zeeman sub-level
K$\left|2\right\rangle $ while the Li atoms remain in the Li$\left|1\right\rangle $
state throughout the experiment. On the time scale of our measurements,
the interactions between these atoms, characterized by the $s$-wave
scattering length $a_{12}$$\approx$$a_{{\rm bg}}$, can be neglected.
We apply a $\pi/2$ rf-pulse (typically 10 $\mu$s-long) to prepare
the K atoms in an equal superposition of the K$\left|3\right\rangle $
and K$\left|2\right\rangle $ states. After a variable interaction
time $\tau$, we apply a second $\pi/2$ rf-pulse before determining
the numbers $N_{2}$ and $N_{3}$ of atoms in the K$\left|2\right\rangle $
and K$\left|3\right\rangle $ states using absorption imaging \cite{MCSupMat}.
To decrease the sensitivity to the magnetic field noise and to the
inhomogeneities in the atom densities, we perform a spin echo by splitting
the interaction time into two equal halves separated by a $\pi$ rf-pulse.

\begin{figure}[t]
\includegraphics[scale=0.38]{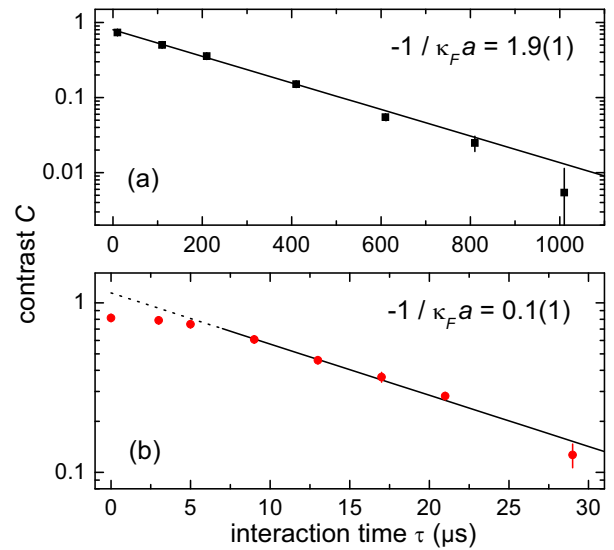}\caption{\label{fig:2}Contrast $C$ as a function of interaction time $\tau$.
In (a), we show results for moderately attractive interspecies interactions
($-1/\kappa_{F}a$=1.9), corresponding to Fig.\ \ref{fig:1}. In
(b), we probe the system in the strongly interacting regime ($-1/\kappa_{F}a$=0.1)
for $T$=$0.20\,\epsilon_{F}/k_{B}$ by rapidly shifting the interaction
parameter from 2.2 to 0.1 during the interaction time. The solid 
lines are exponential fits to the points with $\tau$>7 $\mu$s. The 
dotted line is an extrapolation to $\tau$=0.}
\end{figure}

Shifting the phase of the rf oscillator by $\phi$ between the $\pi$
and the second $\pi/2$ pulses causes a sinusoidal variation in the
fraction $f$=$N_3$/($N_2$+$N_3$) of the K atoms transferred
to K$\left|3\right\rangle $, as shown in Fig.\ \ref{fig:1}. We
quantify the coherence of the state of the K atoms by the contrast
$C$=($f_{{\rm max}}$$-$$f_{{\rm min}}$)/($f_{{\rm max}}$+$f_{{\rm min}}$) 
of these oscillations. The interaction of the K atoms with the Li
cloud causes an exponential decrease in the observed contrast with
increasing interaction time $\tau$, as shown in Fig.\ \ref{fig:2}(a).
The interaction also shifts the rf transition frequency and decreases
the rf coupling between the K$\left|2\right\rangle $ and K$\left|3\right\rangle $
states \cite{Kohstall2012mac}, which we account for by adjusting
the rf frequency and the duration of our rf pulses. In this way, we
measure the decoherence of K atoms for $-1/\kappa_{F}a$<$-$0.8 and
$-1/\kappa_{F}a$>1.4. Near the center of the resonance, the fast
loss of contrast during the rf pulses limits the applicability of
this method.

To measure the decoherence of K in the strongly interacting regime,
we use laser light to rapidly displace our magnetic FR \cite{Bauer2009coam,Bauer2009coa}.
Optical control of our FR allows us to apply the rf pulses 
away from the FR and then rapidly bring the atoms into resonance
for the duration of the interaction time $\tau$ \cite{B0shiftNote}. This method circumvents
the loss of contrast during the rf pulses and allows us to probe the
K decoherence across the full range of interaction parameters%
. The displacement of our FR arises from the laser-induced differential
AC Stark shift between the free-atom level and the molecular state
involved in the FR. The 
AC Stark shift is induced by the 1064-nm trapping light, as we investigated 
in Ref.\ \cite{Jag2014ooa}. Although the differential shift here 
amounts to only 10\% of the total trapping potential, using a high-intensity 
beam with up to 65 kW/cm$^{2}$%
, we can displace $B_{0}$ by up to 40 mG%
{} in less than 200 ns -- all while preserving the harmonic trapping
potential \cite{MCSupMat}. This displacement corresponds to a change
in the interaction parameter of up to $\pm$2.1 
on a timescale of 0.05 $\tau_{F}$, where $\tau_{F}$=$\hbar/\epsilon_{F}$$\approx$4.5 
$\mu$s is the Fermi time.

In Fig.\ \ref{fig:2}(b), we show the dependence of the contrast
$C$ on the interaction time $\tau$ near the center of our FR. The
contrast starts to decay after an initial delay of approximately $\tau_{F}$.
This delay can be explained in terms of quantum evolution of the system
with an interaction energy bounded from above by $\epsilon_{F}$ \cite{Knap2012tdi}.
For $\tau$>1.6$\tau_{F}$$\approx$7 $\mu$s, the decrease in contrast
is well-described by an exponential decay. The fitted decoherence
rate $\gamma_{{\rm coh}}$=0.28(2)$\tau_{F}^{-1}$ is comparable
to the inverse Fermi time, indicating that our experiment cannot be described using the Fermi liquid 
picture, which assumes long-lived quasiparticles \cite{Baym1991lfl}.

\begin{figure}
\includegraphics[scale=1.11]{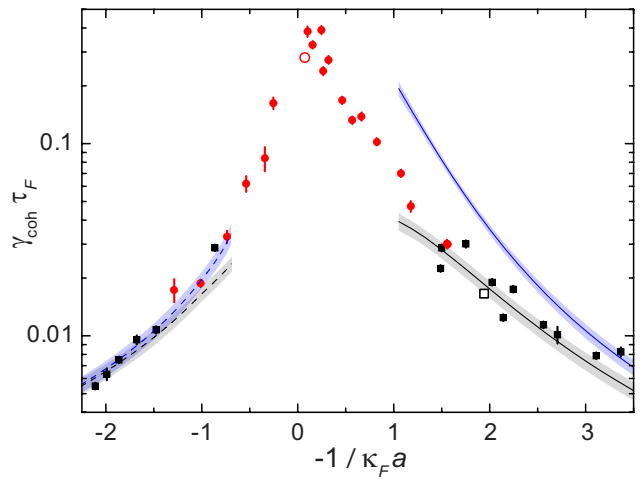}\caption{\label{fig:3}Decoherence rate of K in a Li Fermi sea
as a function of the interaction parameter for an average temperature
$T$=0.16 $\epsilon_{F}/k_{B}$ (see text). The measurements with (without)
rapid shifting of the FR are shown as the red circles (black squares).
The measurements from Fig.\ \ref{fig:2} are indicated by open symbols.
The solid upper (blue) and lower (black) lines correspond to the prediction of the 
Fermi liquid theory with and without medium corrections, respectively. 
The dashed lines incorporate corrections due to decay to Feshbach 
molecules. The shaded areas show the $1\sigma$ effect of the experimental 
uncertainties on the theoretical predictions.}
\end{figure}

In Fig.\ \ref{fig:3} we show the dependence of $\gamma_{{\rm coh}}$
extracted using the same method on the interaction parameter. We present
data with two decades of dynamic range and demonstrate a dramatic
resonant enhancement of the decoherence rate, reaching values up to
$0.4\,\tau_{F}^{-1}$. The data do not exhibit any clear dependence
on $\bar{n}_{{\rm K}}/\bar{n}_{{\rm Li}}$ across the full range 0.17$\leq$$\bar{n}_{{\rm K}}/\bar{n}_{{\rm Li}}$$\leq$0.43.
In addition to the statistical errors indicated 
by the error bars, the data are subject to variations of $k_{B}T/\epsilon_{F}$,
$\kappa_{F}R^{*}$ and $\bar{n}_{{\rm K}}/\bar{n}_{{\rm Li}}$ with
standard deviations of $0.01$, $0.02$ and $0.07$ about their mean
values of $0.16$, $0.93$ and $0.27$, respectively. The calibration
of the Li atom number introduces a 6\% systematic uncertainty in $\epsilon_{F}$
and $\tau_{F}$, as well as a corresponding 3\% uncertainty in $\kappa_{F}$.
Our total error budget includes further 3\% systematic errors in $a$
and $R^{*}$ arising from the uncertainty in $\Delta B$%
, and a $\pm0.05$%
{} error in $1/\kappa_{F}a$ resulting from an uncertainty in the determination
of $B_{0}$ of $\pm$1 mG \cite{MCSupMat}.

For weak to moderate interactions, there are well-defined K quasiparticles, and we now show that 
the evolution of the contrast $C$ on timescales much longer than $\tau_{F}$ can be related to the mean quasiparticle 
scattering rate $\gamma_{s}$. Each scattering event provides which-way 
information that distinguishes between the two paths in the interferometer 
in Fig.\ \ref{fig:1} and thus erases the interference effect. At 
any given time, the interaction affects only one of the two paths, 
decreasing the probability for the system to stay in this path at 
the rate $\gamma_{s}$. Since our signal arises from the interference 
of the amplitudes in the two interferometer paths, we expect the interaction 
to lead to a decrease of the observed contrast at the rate $\gamma_{s}/2$.

From Fermi liquid theory, the scattering rate $\gamma_{p_1}$ 
of a K quasiparticle with momentum ${\bf p}_{1}$ is given by \cite{Baym1991lfl} 
\begin{equation}
\gamma_{p_1}=\iint\! 
d\check p_2
d\Omega\frac{m_rp_r}{4\pi^2}
|\mathcal{T}|^{2}[f_{p_2}^{\text{Li}}(1-f_{p_3}^{\text{K}}-f_{p_4}^{\text{Li}})+f_{p_3}^{\text{K}}f_{p_4}^{\text{Li}}].
\label{Lifetime2}
\end{equation}
 Here $\mathcal{T}$ is the scattering matrix for the scattering of K atoms with Li 
atoms with momenta ${\mathbf p}_1$ and ${\mathbf p}_2$ respectively to 
momenta ${\mathbf p}_3$ and ${\mathbf p}_4$. We have defined $d\check p_2$$=$$d^3p_2/(2\pi)^3$, and 
$\Omega$ is the solid angle for the direction of the outgoing relative momentum. The distribution functions are 
$f_{p}^{\text{Li/K}}$$=$$[ e^{\beta ( E_{p}^{\text{Li/K}} - \mu_{\text{Li/K}}) } + 1 ]^{-1}$ 
with the chemical potentials $\mu_{\text{Li/K}}$ for the Li /K atoms respectively. The dominant 
medium effects can be shown to enter in the scattering matrix 
$\mathcal{T}$ via ladder diagrams, whereas the quasiparticles can be assumed to have the ideal gas energy dispersion
$E_{p}^{\text{K/Li}}$$=$$p^{2}/2m_{{\text{K/Li}}}$~\cite{Bruun2005vat,Enss2012qct}.
The details of the calculation of $\gamma_{p_1}$ are described in ~\cite{Christensen2014qsr}.
In addition, we account for the reduced quasiparticles residue $Z$ by multiplying the collision rate by $Z$ calculated from the ladder approximation \cite{Massignan2014pdm}.
To obtain the mean scattering rate $\gamma_{s}$, we calculate the thermal average $\gamma_s$$=$$\int d\check p f_p^{\text{K}}\gamma_p$.
To include the effects of the trap, we use effective Fermi energies, 
which are obtained by averaging the local Fermi energy over the density of the
K atoms in the trap. 
This approach is justified since the K atoms only probe a small region of the Li 
gas, and because the momentum distribution of the K atoms is nearly classical.

On the repulsive side of the FR, we need to consider additional effects
arising from the decay of the atoms into the molecular state that
underlies our FR. The rate $\Gamma$ of this process was calculated
and confirmed by measurements in Ref.\ \cite{Kohstall2012mac}, reaching
values as high as 0.02 $\tau_{F}^{-1}$ close to resonance. Since
the decay to molecules provides which-way information, it will contribute
at least $\Gamma/2$ to the measured decoherence rate. The decay also
releases energy and creates holes in the Li Fermi sea, increasing
the value of $k_{B}T/\epsilon_{F}$ during our measurement to $0.20\left(1\right)$
\cite{MCSupMat}. We include both of these effects in the calculations.

In Fig.\ \ref{fig:3}, we plot as solid/dashed lines the calculated decoherence rate $\gamma_s/2$ 
on the attractive/repulsive side of the FR for $T$=0.16 $\epsilon_F/k_B$ (attractive side) and $T$=0.20 $\epsilon_F/k_B$ (repulsive side). The lower 
lines are obtained by using the vacuum scattering matrix 
$\mathcal{T}_{\text{vac}}$~\cite{Christensen2014qsr} in (\ref{Lifetime2}), whereas the upper lines are obtained 
by using a $\mathcal{T}$ matrix which includes medium effects using the ladder
approximation. The calculated decoherence rate 
agrees with the experimental values very well for $-1/k_Fa$$\gtrsim$1.5 and for $-1/k_Fa$$\lesssim$$-$1.
This gives strong evidence that the observed decoherence indeed is due to quasiparticle collisions. 
The significant asymmetry 
of the decoherence rate around $1/k_Fa=0$ arises from the narrow nature of the FR \cite{Christensen2014qsr}. 
The calculated decoherence rate is 
larger when medium effects are included in the $\mathcal{T}$ matrix. This is due to pair correlations, which can increase the collisional cross section 
significantly~\cite{Christensen2014qsr}. We see that the inclusion of these medium effects on the scattering matrix reduces the agreement 
with the experimental values on the attractive side, whereas it improves the agreement on the repulsive side. 
We speculate that this intriguing result arises from effects such as shifts in energy and effective mass that are beyond the present theoretical approach.
For stronger interactions, the calculation does not fit the experiment, which is 
expected since there are no well-defined quasiparticles in the unitarity regime~\cite{Kohstall2012mac}.
Our model agrees with the observed absence of a dependence of $\gamma_{{\rm coh}}$ on $\bar{n}_{{\rm K}}/\bar{n}_{{\rm Li}}$ since
the K cloud is close to the classical regime
where $f_{p_{3}}^{{\rm K}}$$\ll$$1$ and the momentum distribution of the K atoms is solely determined by the temperature.

\begin{figure}
\includegraphics[scale=0.37]{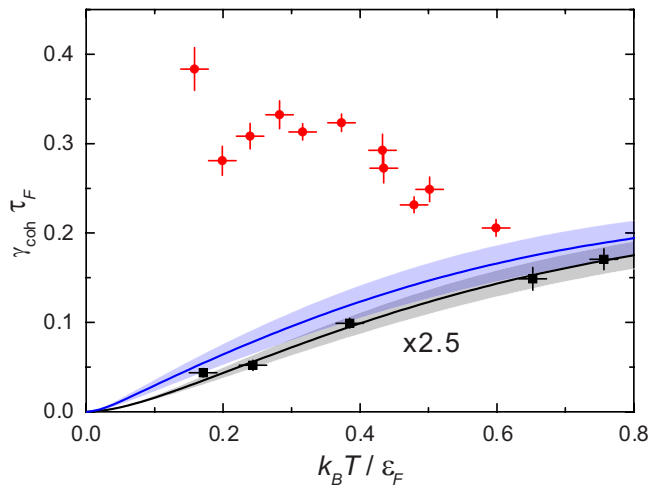}\caption{\label{fig:4}Decoherence rate of K in a fermionic Li cloud as
a function of temperature. The data for $-1/\kappa_{F}a$=0.1, $\kappa_{F}R^{*}$=0.94,
$\bar{n}_{{\rm K}}/\bar{n}_{{\rm Li}}$=0.2 ($-1/\kappa_{F}a$=2.3,
$\kappa_{F}R^{*}$=0.89, $\bar{n}_{{\rm K}}/\bar{n}_{{\rm Li}}$=0.3)
measured with (without) rapid shifting of the FR is shown as red circles
(black squares). The solid blue and black lines correspond to the
predictions of the Fermi liquid theory for $-1/\kappa_{F}a$=2.3 with
and without medium corrections, respectively. The shaded areas show
the $1\sigma$ effect of the experimental uncertainties on the theoretical
predictions.}
\end{figure}

Further insight into the nature of the observed decoherence can be
gained by varying the temperature of our atom mixture, which we accomplish
by changing the endpoint of our evaporative cooling. We show the dependence
of the measured decoherence rate on temperature in Fig.\ \ref{fig:4}.
In addition to the statistical errors shown by the error bars, the
data are subject to small variations of $-1/\kappa_{F}a$, $\kappa_{F}R^{*}$
and $\bar{n}_{{\rm K}}/\bar{n}_{{\rm Li}}$ with standard deviations
of 0.05, 0.03 and 0.1, respectively. Our total error budget also 
includes the above-mentioned systematic uncertainties in $\epsilon_{F}$, 
$\kappa_{F}$, $a$ and $R^{*}$.

Away from the FR, the measured decoherence rates are in very good
agreement with the predictions of the Fermi liquid theory. Including
medium effects in the scattering matrix leads to an overestimate of the collision rate on the attractive side of the FR. 
The linear dependence of $\gamma_{{\rm coh}}$ 
on temperature in this regime arises from the high relative mass of 
the K atoms, causing the Li-K scattering to resemble scattering by 
fixed impurities. This is similar to the situation in metals where 
the scattering of electrons by fixed nuclei gives rise to the well-known 
linear dependence of the nuclear decoherence rates on temperature 
\cite{Korringa1950nmr}. The red circles in Fig.\ \ref{fig:4} represent the measurements of the
decoherence rate for resonant interactions.
The rates obtained in this regime are
more than an order of magnitude higher than the off-resonant rates,
and do not increase with temperature. 

In conclusion, we established that for weak to moderate interaction strengths, the decoherence of K in a Li Fermi sea is dominated by quasiparticle scattering. 
Our observations for strong interactions cannot be explained by quasiparticle scattering and indicate a finite decoherence rate at zero temperature. 
This offers an exciting opportunity to explore the many-body quantum dynamics of an impurity submerged in a Fermi sea.

We thank F. Schreck, C. Kohstall, I. Fritsche, P. Jegli\v{c}, Y. Ohashi and, especially,
M. Parish, J. Levinsen and M. Baranov for helpful discussions. 
We thank P. Massignan for sending us data for the quasiparticle residue and decay.
We acknowledge funding by the 
Austrian Science Fund FWF within the SFB FoQuS (F40-P04) and support
of the Villum Foundation via Grant No. VKR023163.

\bibliographystyle{apsrev}
\bibliography{BIB_PolCoh,ultracold}

\begin{thebibliography}{32}
\bibitem[{B0s()}]{B0shiftNote}
\bibinfo{note}{For measurements on the attractive/repulsive side of the FR, we
  shift $B_0$ upwards/downwards. For measurements near the resonance, we verify
  that the direction of the shift of $B_0$ does not affect the result.}
\end{thebibliography}

\end{document}